\documentclass{llncs}
\usepackage{makeidx}
\usepackage{amssymb,amscd}
\usepackage{graphicx}
\usepackage{float}

\begin{document}

\title{Text Categorization via Similarity Search}

\subtitle{An Efficient and Effective Novel Algorithm\thanks{This work has been partially supported by a 2012 NSERC Canada Graduate Scholarship and a 2013 Ontario Graduate Scholarship (Hubert Haoyang Duan), 2012--2017 NSERC Discovery Grant {\em ``New set-theoretic tools for statistical learning''} (Vladimir Pestov), and the 2012 Mitacs Globalink Program (Varun Singla).}}

\author{Hubert Haoyang Duan\inst{1} \and Vladimir Pestov\inst{1} \and Varun Singla\inst{2}}

\institute{
University of Ottawa, Ottawa, Ontario K1N 6N5, Canada\\
\email{\{hduan065,vpest283\}@uottawa.ca}
\and
Indian Institute of Technology Delhi, Hauz Khas, New Delhi-110 016, India\\
\email{ee5080429@ee.iitd.ac.in}
}

\maketitle

\begin{abstract}
  We present a supervised learning algorithm for text categorization which has brought the team of authors the 2nd place in the text categorization division of the 2012 Cybersecurity Data Mining Competition (CDMC'2012) and a 3rd prize overall. The algorithm is quite different from existing approaches in that it is based on similarity search in the metric space of measure distributions on the dictionary. At the preprocessing stage, given a labeled learning sample of texts, we associate to every class label (document category) a 
  point in the space of question. Unlike it is usual in clustering, this point is not a centroid of the category but rather an outlier, 
  a uniform measure distribution on a selection of domain-specific words. At the execution stage, an unlabeled text is assigned a text category as defined by the closest labeled neighbour to the point representing the frequency distribution of the words in the text.
  The algorithm is both effective and efficient, as further confirmed by experiments on the Reuters 21578 dataset.

\end{abstract}


\section{Introduction}

The amount of texts readily available in the world is growing at an astonishing rate; classifying these texts through machine learning techniques, promptly and without much human intervention, has thus become an important problem in data mining. Much research, in the field of supervised learning, has been done to find accurate algorithms to classify documents in a dataset to their appropriate categories, e.g. see \cite{sebsurvey} or \cite{aasreport} for a detailed survey of text categorization.

The most widely used model for text categorization is the Vector Space Model (VSM) \cite{vsm}. Under this model, a data dictionary $\mathcal T$ consisting of unique words across the documents in the dataset is constructed. The documents are represented by real-valued vectors in the space ${\mathbb{R}}^{\mathcal T}$ with dimension equaling to the size of the dictionary. Given $t\in{\mathcal T}$, the $t$-th coordinate of a vector is the relative frequency of the word $t$ in a given document. When some of the documents' actual class labels are known and used for training, many well-known classifiers in supervised machine learning, such as SVM \cite{svm}, $k$-NN \cite{knn}, and Random Forest \cite{breiman}, can then be applied to categorize documents.

Text categorization decidedly comes across as a problem of detecting
similarities between a given text and a collection of texts of a
particular type. Although distance-based learning rules for text
categorization, such as the $k$-nearest neighbour classifier, e.g.
\cite{knntext}, are not new, they are currently based on the entire feature space, while any dimension reduction steps are done independently
beforehand \cite{sebsurvey}. 

We aim to fill this gap by suggesting a novel supervised learning
algorithm for text categorization, called the Domain-Specific
classifier. It discovers specific words for each category, or domain, of documents
in training and classifies based on similarity searches in the space
of word frequency distributions supported on the respective
domain-specific words.

For each class label, $j=1,2,\ldots,k$, our algorithm extracts class, or domain, specific words from labeled training documents, that is, words that appear in the class $j$ more frequently than in all the other document classes combined (modulo a given threshold). 
Now a given unlabeled document is assigned a label $j$ if the normalized frequency of domain-specific words for $j$ in the document is higher than for any other label. 

To see that this classifier is indeed similarity search based, let $x_j\in {\mathbb R}^{\mathcal T}$ be a binary vector whose $t$-th coordinate is $1$ if and only if $t$ is domain-specific to $j$, and 0 otherwise. Normalize $x_j$ according to the $\ell^p$ distance, and let a document be represented by a vector $w\in{\mathbb R}^{\mathcal T}$. Then the label assigned to $w$ is that of the closest neighbour to $w$ among $x_1,x_2,\ldots,x_k$ with regard to the simplest similarity measure, the inner product on ${\mathbb R}^{\mathcal T}$. In other words, we seek to maximize the value of $\langle w,x_j\rangle$ over $j=1,2,\ldots,k$. Notice that the well-known cosine similarity measure, cf. e.g. \cite{sebsurvey}, corresponds to the special case $p = 2$.

This algorithm was first used in the 3rd Cybersecurity Data Mining Competition (CDMC 2012) to notable success, as the team of authors placed second in the text categorization challenge, and first in classification accuracy \cite{cdmc2012}. In addition, the classification performance of the algorithm was validated on a sub-collection of the popular Reuters 21578 dataset \cite{reuters}, consisting of single-category documents from the top eight most frequent categories, with the standard ``modApt\'e'' training/testing split. In terms of accuracy, our classifier performs slightly better than SVM with a linear kernel, and is significantly faster.

This paper is organized as follows. Section \ref{backg} surveys common feature selection and extraction methods and classifiers considered in the text categorization literature. Section \ref{domainclass} explains the new Domain-Specific classifier in detail and casts it as a similarity search problem. Section \ref{discuss} discusses results from the CDMC 2012 Data Mining competition and experiments from the competition and on the Reuters 21578 dataset. Finally, Section \ref{conclude} concludes the paper with some discussion and directions for future work.

\section{Related Work}\label{backg}

In this section, we describe the VSM model and provide a brief survey on widely known methods for text categorization. 

From this section onwards, following notation similar to \cite{sebsurvey}, we let $\mathcal{D} = \{d_1,d_2,\ldots,d_n\}$ denote the dataset of documents, with size $n = |\mathcal{D}|$, and $\mathcal{T} = \{t_1,t_2,\ldots,t_m\}$ the data dictionary of all unique words from documents in $\mathcal{D}$, with size $m = |\mathcal{T}|$. Given a document $d$ and a word $t\in \mathcal{T}$, $|d|$ denotes the number of words in $d$ and $t\in d$ indicates that the word $t$ is found in $d$.

\subsection{Vector Space Model}

The Vector Space Model (VSM) \cite{vsm} is the most common model for document representation in text categorization. According to \cite{aasreport}, there is usually a standard preprocessing step for the documents in $\mathcal{D}$, where all alphabets are converted to lowercase and all stop words, such as articles and prepositions, are removed. Sometimes, a stemming algorithm, such as the widely used Porter stemmer \cite{porter}, is applied to remove suffices of words (e.g. the word ``connection'' $\to$ ``connect'').

In VSM, the data dictionary, consisting of all unique words that appear in at least one document in $\mathcal{D}$, is first constructed. Sometimes, $n$-grams, which are phrases of words, are also included in the dictionary; however, the benefit of these additional phrases is still up for debate \cite{sebsurvey}. Given the data dictionary, each document can be represented as a vector in the real-valued vector space with dimension equaling the size of the dictionary. Two common methods for associating a document to a vector are explained below.

The simplest method assigns to a document $d$ the vector consisting of the relative term frequencies for $d$, see e.g. \cite{greecesurvey}. The second, known as the $tf$-$idf$ method, assigns $d$ to the vector consisting of the products of term and inverse document frequencies \cite{tfidf}. Mathematically speaking, a document $d$ is mapped to a real-valued vector of length $m$: $d \longmapsto (w_1,w_2,\ldots,w_m)\in \bbbr^m$, for
\begin{equation}
  \label{eq:frequency}
w_i = \frac{c(t_i,d)}{|d|} \quad\mbox{(frequency method)}
\end{equation}
or
\begin{equation}
  w_i = c(t_i,d)\log\left(\frac{n}{|\{d\in \mathcal{D}: t_i \in d\}|}\right) \quad\mbox{($tf$-$idf$ method)},
\end{equation}
where $c(t_i,d)$ denotes the number of times the word $t_i$ appears in $d$. Other representations include binary and entropy weightings and the normalized $tf$-$idf$ method \cite{aasreport}.

Once the documents are represented as vectors, the dataset can be interpreted as a data matrix $M$ of size $(n\times m)$. However, a main challenge for text categorization is that the size of the data dictionary is usually immense so the data matrix is extremely high dimensional. Dimension reduction techniques must often be applied before classification to reduce complexity \cite{aasreport}.

\subsection{Feature Selection and Extraction Methods}

Due to the potentially large size of the data dictionary, feature selection and extraction methods are often applied to reduce the dimension of the data matrix. Feature selection methods assign to each feature, a word in the data dictionary, a statistical score based on some measure of importance. Only the highest scored features, past some defined threshold, are kept and a lower dimensional data matrix is created from only these features. Some known feature selection methods in text categorization include calculating the document frequency, e.g. \cite{selectionyang}, mutual information \cite{mutual}, and $\chi^2$ statistics, e.g. \cite{chisq}. See e.g. \cite{selection} or \cite{selectionyang} for a thorough study of text feature selection methods.

Feature extraction methods transform the original list of features to a smaller 
list of new features, based on some form of feature dependency. Common well-known feature extraction techniques, as surveyed in \cite{milos}, are Latent Semantic Indexing (LSI) \cite{lsi}, Linear Discriminant Analysis (LDA) \cite{lda}, Partial Least Squares (PLS) \cite{pls}, and Random Projections \cite{randomproj}.

\subsection{Classification Algorithms}

Well-known classifiers that have been applied to text categorization include the $k$-nearest neighbour classifier \cite{knntext}, Support Vector Machines \cite{svmtext}, the Naive Bayes classifier \cite{naivestext}, and decision trees \cite{dttext}. We ask the reader to refer to indicated references, or to survey articles, such as \cite{sebsurvey}, \cite{greecesurvey}, and \cite{aasreport}. The paper \cite{yangexam} provides comparable empirical results on some of these classifiers.

The standard approach in literature for text categorization is that one, or more, feature selection or extraction technique is first applied to a data matrix, since the original data matrix is often extremely high-dimensional. A learning algorithm, independent of the dimension reduction process, is then used for classification \cite{sebsurvey}. The novel approach in this paper is that we consider a new classifier based only on extracted class specific words, which naturally reduces time complexity and the dimension of the dataset. In other words, the Domain-Specific classifier both performs dimension reduction and classifies, in consecutive and dependent steps.

\section{The Domain-Specific Classifier}\label{domainclass}

Our algorithm consists of two distinct stages: extraction of domain-specific words from training samples and classification of documents based on the closest labeled point determined by these domain-specific words. 

\subsection{Preprocessing Stage: Domain-Specific Words}

Fix an alphabet $\Sigma$ and denote $\Sigma^*$ as the set of all possible ``words" formed from $\Sigma$. A document $d$ is then simply an ordered sequence of ``words", $d\in (\Sigma^*)^{|d|}$, and the data dictionary $\mathcal{T}$ is a subset of $\Sigma^*$. Given a set of labeled documents, we can denote it as $\mathcal{D}_\mathrm{lab} = \{(d_1,l_1),(d_2,l_2),\ldots,(d_n,l_n)\}\enspace$, where $d_i$ is a document and $l_i\in\{1,2,3,\ldots,k\}$ is its label, out of a possible $k$ different labels. In addition, we can partition $\mathcal{D}_\mathrm{lab}$ into subsets of documents according to their labels:
\begin{equation}
\mathcal{D}_\mathrm{lab} = \bigcup_{j = 1}^k \mathcal{D}_\mathrm{lab}^j
\end{equation}
where $\mathcal{D}_\mathrm{lab}^j = \{(d,l)\in \mathcal{D}_\mathrm{lab} : l = j\}$ is the set of documents of label $j$. Then, for a particular label $j$ and a word $t\in \mathcal{T}$ in the data dictionary, we denote $f_j(t)$ as the average proportion of times the word $t$ appears in documents with label $j$:
\begin{equation}
f_{j}(t) = \frac{1}{|\mathcal{D}_\mathrm{lab}^j|}\sum_{(d,j)\in\mathcal{D}_\mathrm{lab}^j}  \frac{c(t,d)}{|d|}\enspace.
\end{equation}

Domain-specific words are those words which appear, on average, proportionally more often in one label type of documents in $\mathcal{D}_\mathrm{lab}$ than other types.

\begin{definition}
  Let $\alpha \geq 0$. A word $t\in\mathcal{T}$ in the data dictionary is {\em domain} (or {\em class}) {\em $j$ specific} if
\begin{equation}
f_j(t) > \alpha \sum_{j' \neq j} f_{j'}(t).
\end{equation}
\end{definition}

This definition of domain-specific words depends on the parameter $\alpha$ and hence, so does the Domain-Specific classifier. As $\alpha$ increases from $0$, the number of domain-specific words for each class label decreases; as a result, $\alpha$ can be thought of as a threshold parameter, and an optimal choice for $\alpha$ is determined through cross-validation using training data.

\subsection{Classification Stage}

Let now $d$ be an unclassified document. We associate to it a vector $w=w_d\in {\mathbb R}^{\mathcal T}$ (a relative frequency distribution of words) as in Eq. (\ref{eq:frequency}), that is, for every $t\in {\mathcal T}$,
\begin{equation}
  \label{eq:w}
  w(t) =  \frac{c(t,d)}{|d|}.\end{equation}

Let $j$ be a label. Denote $CS_j=CS_{j,\alpha}$ the set of domain-specific words to $j$. Define the total relative frequency of domain $j$ specific words found in $d$:
\begin{equation}
  w[CS_j] =\sum_{t\in CS_j} w(t)=\frac{1}{\vert d\vert}\sum_{t\in CS_j} c(t,d).
\end{equation}
The classifier assigns to $d$ the label $j$ for which the following ratio is the highest: 
\begin{equation}
  \label{eq:classifier}
  j=\mbox{argmax}_i\,\frac{w[CS_i]}{\vert CS_i\vert^{1/p}}.
\end{equation} 
Here, $p\in\left(0,\infty\right]$ is a parameter, which normalizes a certain measure with regard to the $\ell^p$ distance, cf. below in Section \ref{spacemeasure}.

\subsection{Space of Positive Measures on the Dictionary}\label{spacemeasure}

A (positive) measure on a finite set $\mathcal T$ is simply an assignment $t\mapsto w(t)$ to every $t\in {\mathcal T}$ of a non-negative number $w(t)$; a probability measure also satisfies $\sum_{t\in {\mathcal T}}w(t)=1$. Denote $M(\mathcal{T})$ the set of all positive measures on $\mathcal{T}$.

Fix a parameter $p\in\left(0,\infty\right]$. The following is a positive measure on $\mathcal{T}$:
\begin{equation}
  x_j(t) = \left\{\begin{array}{cl}
  \frac{1}{\vert CS_j\vert^{1/p}},&\mbox{ if }t\in CS_j,\\
  0,&\mbox{ otherwise.}
  \end{array}\right.
  \end{equation}
If $p = 1$, we obtain a probability measure uniformly supported on the set of domain $j$ specific words. In general, values of $p \in\left(0,\infty\right]$ correspond to different normalizations of the uniform measure supported on these words, according to the $\ell^p$ distance. (The case when $p = \infty$, that is, the $\ell^\infty$ distance, corresponds to non-normalized uniform measure.)

Among the similarity measures on $M(\mathcal T)$, we single out the standard inner product
\begin{equation}\langle w,v\rangle = \sum_t w_tv_t.\end{equation}
Notice that for every $w\in M(\mathcal T)$ and each $j$,
\begin{equation}\langle w,x_j\rangle = \frac{w[CS_j]}{\vert CS_j\vert^{1/p}},\end{equation}
and for this reason, the classification algorithm (\ref{eq:classifier}) can be rewritten as follows:
\begin{equation}
  \label{eq:classifier2}
  j=\mbox{argmax}_i\,\langle w,x_i\rangle.
  \end{equation}
Our classifier is based on finding the closest point $x_i$ to the input point $w$ in the sense of the simplest similarity measure, the inner product.

The similarity workload is a triple $({\mathbb U},S,{\mathbb X})$, consisting of the domain ${\mathbb U}=M(\mathcal T)$, the similarity measure $S(w,v)=\langle w,v\rangle$ equal to the standard inner product (a rather common choice in the problems of statistical learning, cf. \cite{SS}), and the dataset ${\mathbb X}=\{x_1,x_2,\ldots,x_k\}$ of normalized uniform measures corresponding to the text categories and domain-specific words extracted at the preprocessing stage.

Note that the well-known cosine similarity measure arises in the special case when the normalizing parameter is $p = 2$; hence, it is not necessary to consider this measure separately. Our experiments have shown that different datasets require different normalizing parameters for $x_j$, and that the optimal normalization depends on the sizes of the document categories; Section \ref{conclude} includes a discussion on this topic.

\section{Experiments and Results}\label{discuss}

This section details the experiments and results obtained for the Domain-Specific classifier, in the 2012 Cybersecurity Data Mining Competition and on the Reuters 21578 dataset. All of the programming for this section were done with standard packages in R \cite{r} and with the specialized packages $\texttt{e1071}$ \cite{e1071} and $\texttt{randomForest}$ \cite{randomf}, on a desktop running Windows 7 Enterprise, with a Intel i5 3.10 GHz processor and 4GB of RAM.

\subsection{The 2012 Cybersecurity Data Mining Competition}

The 3rd Cybersecurity Data Mining Competition (CDMC 2012) \cite{cdmc2012}, associated with the 19th International Conference on Neural Information Processing (ICONIP 2012) in Doha, Qatar from November 12 - 15, 2012 \cite{iconip}, included three supervised classification tasks: electronic news (e-News) text categorization, intrusion detection, and handwriting recognition. 

The Domain-Specific classifier was first developed by the team of authors for the e-News text categorization challenge, which required classifying news documents to five topics: business, entertainment, sports, technology, and travel. The documents were collected from eight online news sources. The words in these documents were obfuscated, and all punctuations and stop words removed. Here is a sample scrambled text document paragraph from the competition:

\begin{quote}
  \scriptsize
 HUJR Xj gjXZMUXe fAJjAeK UO jwXeA URSek UYjmX xjI K SeeW eOWrjJeeR ZARWZDek UAk WDjkmzXZMe KXR UA eRReAXZUr BmeRXZjA RZAze zjOWUAZeR XjkUJ OmRX UzzjOWrZRx OjDe IZXx weIeD WejWre uxe 
 OjRX RmzzeRRwmr RXUDXmWR OmRX 
\end{quote}

In total, 1063 e-News documents for training, each labeled as one of the $k = 5$ topics, were given for the goal of classifying 456 documents.  

\begin{table}{\scriptsize
\begin{center}
\caption{Information on the dataset size for e-News classification task.}\label{datasetinfo}
\begin{tabular}{|c|l|c|}
\hline
Label $j$&Topic & \# of documents\\
\hline
\hline
1&Business & 205\\
2&Entertainment & 215\\
3&Sport &193\\
4&Technology & 223\\
5&Travel & 227\\
\hline
&Training set & 1063\\
\hline
\hline
&Classification task& 456\\
\hline
\end{tabular}
\end{center}}
\end{table}

\subsection{Competition Experiments}\label{cdmcexp}

After a pre-processing step, where all document words of length less or equal to 3 were removed, a data dictionary of all unique words from the training and classification documents, consisting of $m = 55822$ words, was constructed. The 1063 labeled documents were converted to vectors of length $m = 55822$, 
according to the Vector Space Model. Then, $5$-fold cross-validation on the training dataset was performed to test the performance of the classifier. 

For comparison purposes, the Support Vector Machines (SVM) classifier \cite{svm}, using the Gaussian Radial Basis (GRB) and linear kernels with cost 10, and the Random Forest classifier \cite{breiman}, using $50$ trees, were also tested. The performance measures considered were classification accuracy and the F-Measure (F1) \cite{aasreport}. See Table \ref{acctable}, where the computation times for both the training and predicting stages are also indicated.
\begin{table}{\scriptsize
\begin{center}
\caption{Classification performance of the Domain-Specific classifier (DSC) through $5$-fold cross validation on the training set, compared to SVM with the Gaussian Radial Basis (GRB) and linear kernels and Random Forest (RF).}
\label{acctable}
\begin{tabular}{|lc||c|c|c|c|c|c|c|c||c|c||c|}
\hline
&& DSC & DSC &DSC &DSC &DSC &DSC &DSC &DSC&SVM&SVM&RF\\
&$\alpha$ &0& 0.25 & 0.5 & 0.75 &  1 & {\bf 2} & 5 & 10 & GRB &linear& \\
\hline
Accuracy && 0.575& 0.854    &  0.887   &   0.896   &  0.896 & {\bf 0.915} & 0.901 & 0.882 & 0.613 & 0.821 & 0.882\\
\hline
F1 Business&& 0.491&   0.765  & 0.839   &   0.833   &  0.765 & 0.794 & 0.774&  {\bf 0.868}& 0.576 & 0.740 & 0.853\\

F1 Entertainment&& 0.400&   0.925  & 0.915   &   0.896   &  0.928 & {\bf 0.949} & 0.926& 0.897 & 0.530 &0.784  & 0.845\\

F1 Sport&& 0.899&   0.962  &   0.946 &   1.000   & 0.987  & {\bf 1.000} &0.974 & 0.944 & 0.847 & 0.886 & 0.943\\

F1 Technology&& 0.548&   0.767  &  0.850  &   0.825   & 0.871  & {\bf 0.891} & 0.891 &  0.848& 0.597 &  0.785& 0.842\\
F1 Travel&& 0.600&  0.883   & 0.892   &    {\bf 0.947}  & 0.916  & 0.920 &0.911 & 0.867 & 0.641 & 0.896 & 0.917\\

\hline
\end{tabular}
\vskip .2cm

Computational Time

\begin{tabular}{|c||c||c|c||c|}
\hline
& DSC & SVM& SVM& RF\\
& & GRB & linear & \\
\hline
Training stage & {\bf 1.5 secs}&3.99 mins&3.25 mins&3.96 mins\\
Predicting stage & 0.6 secs &14.3 secs &16.9 secs&{\bf 0.4 secs}\\
\hline
\end{tabular}
\end{center}}
\end{table}

The choice $\alpha = 2$ resulted in the best accuracy of 0.915 for the Domain-Specific classifier, and the optimal normalizing parameter was $p = 1$, corresponding to the choice of $x_1,x_2,\ldots,x_k$ normalized as probability measures uniformly supported on the domain-specific words. Consequently, these two values were used for the classification of the 456 documents in the competition. Note that the accuracy score and the F-Measure for 4 out of the 5 categories, for $\alpha = 2$, were higher than the respective scores obtained with SVM with the GRB and the linear kernels and Random Forest. Experiments had also shown that the Domain-Specific classifier was extremely fast and efficient, since distance calculations are only based on domain-specific words, not the entire data dictionary. As a result, no dimension reduction technique prior to classification was required. 

\subsection{Competition Results}

The submissions for the three tasks for the 2012 Cybersecurity Data Mining Competition \cite{cdmc2012} were strictly evaluated based on the F-Measure with respect to each class label to determine the overall rankings. However, the classification accuracy scores for the three tasks were also sent to the participants.

\begin{table}\scriptsize
\begin{center}
\caption{Results of the Domain-Specific classifier for the e-News task.}
\label{results}
\begin{tabular}{|c||c|c|c|c|c||c||c|}
\hline
& Label 1 & Label 2 & Label 3 & Label 4 & Label 5 & Accuracy & Task Ranking \\
\hline
F-Measure &0.847  &     0.943     &  0.991      & 0.805 &0.947 & - & 2nd\\
\hline
Accuracy & - & - & - & - & - & 0.912 & 1st\\
\hline
\end{tabular}
\end{center}
\end{table}

The team of authors finished 1st in pure accuracy and 2nd in the e-News text categorization task with the Domain-Specific classifier, see Table \ref{results}. Overall, the team received 3rd place in the entire competition. 

\subsection{Experiments on the Reuters 21578 Dataset}

The Reuters 21578 dataset, consisting of documents from the Reuters newswire in 1987 and categorized by Reuters Ltd. and Carnegie Group, Inc., is a classical benchmark for text categorization classifiers \cite{reuters}. To further test the effectiveness and efficiency of the Domain-Specific classifier, we considered single-category documents from the top eight most frequent classes (known as the R8 subset) from the Reuters 21578 dataset and divided according to the standard ``modApt\'e'' training/testing split. These documents were downloaded from \cite{r8}.

Table \ref{r8info} provides the category sizes for this dataset. Standard pre-processing of the dataset consisted of removing all stop words and words of length two or less; afterwards, the size of the dictionary of all unique words from the training and testing documents was $m = 22931$ words.

\begin{table}\scriptsize
\begin{center}
\caption{Information on the Reuter 21578 dataset considered.}\label{r8info}
\begin{tabular}{|c|l|c|c|}
\hline
Label $j$&Topic & \# of training documents & \# of testing documents\\
\hline
\hline
1&acq & 1596&696\\
2&crude& 253&121\\
3&earn&2840&1083\\
4&grain& 41&10\\
5&interest& 190&81\\
6&money-fx&206&87\\
7&ship& 108&36\\
8&trade& 251&75\\
\hline
&Total & 5485 & 2189\\
\hline
\end{tabular}
\end{center}
\end{table}

In this case, evaluation of the Domain-Specific classifier, based on the accuracy, F-Measure, and computational time, has shown that $\alpha = 0.45$ and 
$p = \infty$ (non-normalized measures on domain-specific words) were optimal. The SVM, using the linear kernel and a class weight adjustment (2840 divided by the number of documents in each category) to address the varying sizes of the categories, and Random Forest, using 50 trees, classifiers were also tested to compare against our novel algorithm. Table \ref{r8table} provides the classification results obtained by the Domain-Specific classifier at those values, SVM, and Random Forest.

\begin{table}\scriptsize
\begin{center}
\caption{Classification performance of the Domain-Specific classifier (DSC) compared to SVM with the linear kernel and Random Forest (RF) on the Reuters 21578 dataset.}
\label{r8table}
\begin{tabular}{|c||c||c|c|c|c|c|c|c|c|}
\hline
& Accuracy & F1 acq & F1 crude & F1 earn & F1 grain & F1 interest & F1 money-fx & F1 ship & F1 trade\\
\hline\hline
DSC $\alpha = 0.45$ &{\bf 0.952} &{\bf 0.961}&{\bf 0.954}& 0.978&0.800&{\bf 0.857}&{\bf 0.859}&{\bf 0.836}&0.807\\
\hline
SVM linear &0.946& 0.948&0.913&0.970&{\bf 0.900}&0.834&0.844&0.833&{\bf 0.947}\\
\hline
Random Forest &0.926 & 0.917 &0.911&{\bf 0.983}&0.462&0.775&0.556&0.326&0.877\\
\hline
\end{tabular}

\vskip .2cm
Computational Time

\begin{tabular}{|c||c||c|c|}
\hline
& DSC & SVM& RF\\
& & linear & \\
\hline
Training stage &{\bf 6.6 secs}&1.43 hours &54.55 mins\\
Predicting stage & 2.3 secs&55.9 secs& {\bf 1.08 secs}\\
\hline
\end{tabular}
\end{center}
\end{table}

The Domain-Specific classifier performed slightly better than SVM with the linear kernel, and better than Random Forest in terms of accuracy. With respect to the F-Measure, our classifier performed better than SVM for categories with large sizes, and better than Random Forest in 6 of the 8 categories, while SVM had a higher F-Measure on two of the smaller categories, undoubtably due to SVM's class weight adjustment. Computationally, our classifier ran considerably faster than SVM and Random Forest.

\section{Conclusion}\label{conclude}

In this paper, we have introduced a novel text categorization algorithm, the Domain-Specific classifier, based on similarity searches in the space of measures on the data dictionary. The classifier finds domain-specific words for each document category, which appear in this category relatively more often than in the rest of the categories combined, and associates to it a normalized uniform measure supported on the domain-specific words. For an unlabeled document, the classifier assigns to it the category whose associated measure is most similar to the document's vector of relative word frequencies, with respect to the inner product. The cosine similarity measure arises as a special case corresponding to the $\ell^2$ normalization.

Our classifier involves a similarity search problem in a suitably interpreted domain. We believe that this is the right viewpoint with the aim of further improvements. It is worthwhile noting that our algorithm is unrelated to previously used distance-based algorithms (e.g. the $k$-NN classifier \cite{knntext}). The dataset in the similarity workload is completely different, and as a result, unlike most algorithms in text categorization, this classifier does not require any separate dimension reduction step beforehand.

The process of selecting domain-specific words in our algorithm is actually an implicit feature selection method which is class-dependent, something we have not seen before from a classifier in text categorization. For each class, instead of a centroid, we are choosing an outlier, a uniform measure supported on the domain-specific words, which is representative of this class and not of any other class. Not only does each uniform measure lead to a reduction in the dimension of the feature space (as most words are not domain-specific) for similarity calculations, it does so dependent of the class labels, since domain-specific words are chosen relative to all classes.

This algorithm was first developed for the 2012 Cybersecurity Data Mining Competition and brought the team of authors 2nd place in the text categorization challenge, and 1st place in accuracy. This is evidence that our algorithm outperformed many existing text categorization algorithms, as surveyed in Section \ref{backg}. In addition, our algorithm was evaluated on a sub-collection of the Reuters 21578 dataset against two state-of-the-art classifiers, and shown to have a slightly higher classification accuracy than SVM, with a higher F-Measure for the larger categories, and overall performed better than Random Forest. Computationally, our classifier ran significantly faster than either, especially in the training stage.

The normalizing parameter $p$ plays a significant role: it is to account for class imbalance. When there are categories with very few documents, $p = \infty$ should be used to avoid over-emphasizing the smaller categories; and small values of $p$ should be used when the categories have roughly the same number of documents.

For future work, we hope to test the Domain-Specific classifier on biological sequence databases. Other definitions of domain-specific words can be investigated, for instance the one proposed in \cite{discrim}. We would like to experiment with assigning non-uniform measures on the domain-specific words, for instance, by putting weights based on their relative occurrences or on $\alpha$. Finally, we would like to extend the process of selecting domain-specific words to a general classification context, by defining class-specific features relative to the classes and performing classification on only these class-dependent features.

\end{document}